\documentclass[twocolumn,prl,aps,noshowpacs]{revtex4}
\usepackage{psfig,amssymb,graphics,epsf,psfrag}

\begin{document}
\input epsf

\title{Current fluctuations in non-equilibrium diffusive systems: an additivity principle}
\author{T. Bodineau$^{\dag }$ and B. Derrida$^{\ddag}$ }
\affiliation{$\dag$ Laboratoire de Probabilit{\'e}s et Mod{\`e}les Al{\'e}atoires,
CNRS-UMR 7599, Universit{\'e}s Paris VI $\&$ VII,
4 place Jussieu, Case 188, F-75252 Paris, Cedex 05;\\
$^{\ddag}$ Laboratoire de Physique Statistique, Ecole Normale Sup{\'e}rieure,
24 rue Lhomond, 75231 Paris Cedex 05, France  \\
{\rm emails: bodineau@math.jussieu.fr  and derrida@lps.ens.fr}}

\begin{abstract}
We  formulate a simple additivity principle
allowing to calculate the whole distribution of  current fluctuations through a large one dimensional system in
contact with  two 
reservoirs at unequal densities from the knowledge of its first  two cumulants.  
This distribution  (which in general is non-Gaussian) satisfies the Gallavotti-Cohen symmetry  and generalizes  the one predicted recently for the symmetric  simple
exclusion process. The additivity principle  can be used to study more complex diffusive  networks including loops. 

\vspace{+0.2in}
{02.50.-r, 05.40.-a, 05.70 Ln, 82.20-w}
\end{abstract}

\today
 \maketitle


Understanding the  fluctuations of the steady state  current through a
system in contact with two (or more) heat or particle reservoirs is one
of the simplest and most fundamental  problems of non-equilibrium physics
\cite{LLP,BLR,BlB}.
For quantum particles such as (weakly interacting) electrons which satisfy  the Pauli principle, 
 the whole distribution (the full counting statistics) of the number of particles  
transferred  between the two reservoirs during a long time interval is known
 \cite{LLY}  and it  can be calculated by a number of theoretical
 approaches \cite{BlB,BeB,Nazarov,GG}, ranging from the theory of
random matrices \cite{LLY,Be} to the Boltzmann-Langevin semiclassical
description \cite{Nagaev}.

For systems of purely classical interacting particles \cite{LLP,BLR} in contact with two reservoirs the theory is, to our knowledge, less developed.
However,  for a number of
 stochastic  models  of classical interacting particles \cite{deJong,RD,PJSB,DDR}, the cumulants of the current fluctuations  were found to  coincide with those previously known of non-interacting quantum particles.
It is of course an important issue to know under what condition a classical particle system could present the same
distribution of current as in the quantum case. 

For most theoretical approaches developped  in the quantum or in the classical description,
the calculation of the cumulants becomes harder and harder as the degree of the cumulants increases.
The goal of the present letter is to show that for classical stochastic
models, if one postulates a simple additivity principle for the current
fluctuations, the whole distribution of current fluctuations can be calculated from the  knowledge of the first two cumulants of the current.

We consider here a  one dimensional diffusive open system of length $N$ (with $N$ large) in contact, 
at its two ends, with two reservoirs of particles at densities $\rho_a$ and $\rho_b$. 
In the bulk, the system evolves under some conservative stochastic dynamics 
and, at the boundaries, particles are created or annihilated to match the
densities of the reservoirs.

Let $Q_t$ be the integrated current up to time $t$, i.e. the number of particles which 
went through the system during time $t$.
For large $N$, we shall see that the whole distribution of the fluctuations of $Q_t$ 
depends only on two macroscopic parameters $D(\rho)$ and $\sigma(\rho)$ defined as follows:
Suppose that  for $ \rho_a = \rho + \Delta \rho $ and $\rho_b= \rho $ with 
$\Delta \rho$ small, we know that in the steady state Fick's law holds
\begin{eqnarray}
\label{eq: diffusivite}
{\langle Q_t \rangle \over t} = {1 \over N} D(\rho) \  \Delta \rho \, .
\end{eqnarray}
Suppose  that for $\rho_a= \rho_b=\rho$ (in which case $\langle Q_t
\rangle=0$), we also know that for large $t$ the variance is
\begin{eqnarray}
\label{eq: conductivite}
{\langle Q_t^2 \rangle \over t} = {1 \over N} \sigma(\rho) \, .
\end{eqnarray}
The main result of the present paper is that, using a simple additivity principle (\ref{add1},\ref{scaling}), we can  predict {\it all} the cumulants of  $Q_t$ for arbitrary 
$\rho_a$ and $\rho_b$.
If we define the integrals $I_n$ by
$$I_n = \int_{\rho_b}^{\rho_a} D(\rho)\  \sigma(\rho)^{n-1} \  d \rho \,
,$$
the first cumulants of $Q_t$ are given by
\begin{equation}
{\langle Q_t \rangle \over t} = {1 \over N} I_1,
\quad
 {\langle Q_t^2 \rangle - \langle Q_t
\rangle^2
\over t} = {1 \over N} {I_2\over I_1}
\label{Q12}
\end{equation}
\begin{equation}
{\langle Q_t^3 \rangle_{ c} \over t}=  {1 \over N} { 3 (I_3 I_1 -
I_2^2) \over I_1^3}
\label{Q3}
\end{equation}
\begin{equation}
{\langle Q_t^4 \rangle_{ c} \over t}=  {1 \over N} { 3 ( 5 I_4 I_1^2  - 14
I_1 I_2 I_3 + 9  I_2^3  ) \over I_1^5} 
\label{Q4}
\end{equation}
The case $\rho_a= \rho_b$ can be   obtained by letting $\rho_a$ tend to $\rho_b$
in the previous expressions.

More generally,  all the higher cumulants can be obtained from the knowledge 
of $\mu_N$ which characterizes the large $t$ growth of the  
generating function  of $Q_t$
\begin{equation}
 \mu_N (\lambda,\rho_a,\rho_b)  = \lim_{t \to \infty } 
 t^{-1} \ln \langle e^{\lambda Q_t}   \rangle 
 \,  .
\label{mu-def}
\end{equation}
We are going to show that, for large $N$, $\mu_N$ takes
 the following parametric form 
\begin{equation}
\mu_N(\lambda,\rho_a,\rho_b) =  
-   {K  \over N} \left[ \int_{\rho_b}^{\rho_a}  { D(\rho) \ d \rho
\over \sqrt{ 1 + 2 K \sigma(\rho)}} \right]^2   +o\left( {1 \over N } \right)\, ,
\label{mu}
\end{equation}
where $K = K(\lambda,\rho_a,\rho_b)$ is the solution of
\begin{equation}
\lambda= 
\int_{\rho_b}^{\rho_a} d \rho { D(\rho) \over
\sigma(\rho)}\left[{1
\over \sqrt{ 1 + 2 K \sigma(\rho)}} -1 \right] \, .
\label{lambda}
\end{equation}
As $\mu_N =( \lambda \langle Q_t \rangle+ \lambda^2 \langle Q_t^2 \rangle_{ c}/2+\lambda^3 \langle Q_t^3 \rangle_{ c} /6 +... )/t$, 
one  simply needs to expand
(\ref{mu}) and (\ref{lambda})  in powers of $K$ and  to eliminate $K$ to obtain $\mu_N$ as a power series of $\lambda$ and the cumulants such as (\ref{Q12}-\ref{Q4}).

Note that  (\ref{mu}) and (\ref{lambda}) are only valid  for $\rho_a \neq \rho_b$ and in the range  of values of $\lambda$
where $K$ is large enough  for the argument of the square root in the
integrants  not to vanish. We checked that  they  can also be analytically continued to cover the other ranges  of $\lambda$ and the case $\rho_a=\rho_b$.

Our derivation of (\ref{mu}) and (\ref{lambda}) is based on an additivity
principle  that we are  going to formulate now. The probability $P_N(q,\rho_a,\rho_b,t)$ of observing an integrated  current  $ Q_t = q t$   is exponential in $t$ for large $t$
\begin{equation}
P_N(q,\rho_a,\rho_b,t) \sim \exp[ t \;
F_N(q,\rho_a,\rho_b,{\rm contacts}) ] \,,
\label{ld}
\end{equation}
where 
$F_N(q,\rho_a,\rho_b,{\rm contacts})$ depends on the length $N$ of the
system, on $q$,  on the densities $\rho_a$ and $\rho_b$ in the two reservoirs, and
on the nature of the contacts of the system with the two reservoirs.
($F_N$ is negative and vanishes only   when  $q$ takes its most likely value $ \langle Q_t \rangle / t$).
When $N$ is large and $q$ is of order $1/N$, the effect of the contacts becomes negligible and 
asymptotically $F_N(q,\rho_a,\rho_b)$ depends only on $q,\rho_a,\rho_b$, on  the 
length $N$ and on the bulk properties of the system.
We then assume  that, for large $N$ and $q$ of order $1/N$, the large deviation function $F_N(q,\rho_a,\rho_b)$
satisfies the following {\it additivity principle}:
\begin{equation}
F_{N+N'}(q, \rho_a,\rho_b) \simeq \max_\rho \left\{ F_N(q,\rho_a,\rho) +
F_{N'}(q,\rho,\rho_b) \right\} \, .
\label{add1}
\end{equation}
This property simply means that the two subsystems are independent,
except that they try to ajust the density $\rho$ at their contact 
to maximize the following product
\begin{eqnarray*}
P_{N+N'}(q,\rho_a,\rho_b,t) \sim
\max_\rho \left[
P_{N}(q,\rho_a,\rho,t) \ P_{N'}(q,\rho,\rho_b,t)
 \right] \, .
\end{eqnarray*}
We make also the following {\it scaling hypothesis}
\begin{equation}
F_N(q,\rho_a,\rho_b) \simeq  N^{-1} \  G(Nq,\rho_a,\rho_b) \, .
\label{scaling}
\end{equation}
This hypothesis, which is valid  in particular for the symmetric simple
exclusion process,   means that  $\mu_N$ defined by (\ref{mu-def}) is of order $1/N$ for large $N$ (see \cite{DDR}).

If   we write $N=(N+N')x$, i.e. we split a system of macroscopic  unit length into two parts of lengths $x$ and $1-x$, then  (\ref{add1},\ref{scaling})
 lead to
\begin{eqnarray}
G(q,\rho_a,\rho_b ) 
= \max_\rho \left\{  {  G(q x,\rho_a,\rho) \over x} +
{  G(q (1-x),\rho,\rho_b) \over 1 -x} \right\} 
\label{Gad}
\end{eqnarray}
If we keep dividing the system into smaller and smaller pieces and we use 
 that for a piece of small (macroscopic) size $\Delta x$  (i.e. of 
$N \Delta x$ sites) one has   (\ref{eq: diffusivite},
\ref{eq: conductivite},\ref{add1},\ref{scaling})
\begin{eqnarray}
\frac{1}{\Delta x} G( q \Delta x  ,\rho,\rho + \Delta \rho) \simeq 
-   \, \frac{\left[ q \Delta x  +  D(\rho) \, \Delta \rho  \right]^2}{2 \sigma(\rho) \Delta x} \,.
\end{eqnarray}
one finds  a variational form for $G$
\begin{eqnarray}
\label{eq: fonctionnelle}
G(q,\rho_a,\rho_b) = - \min_{\rho(x)}  \left[ \int_{0}^{1} {
\big[ q + D(\rho(x)) \rho'(x) \big]^2  \over 2 \sigma(\rho(x))} \ dx  
\right] 
\end{eqnarray}
where the minimum is over all the functions $\rho(x)$ with boundary conditions
$\rho(0)=\rho_a$  and $ \rho(1)=\rho_b$.

The optimal  $\rho(x)$ in (\ref{eq: fonctionnelle})
satisfies
$$q^2 a'(\rho) - c'(\rho) \left( {d \rho \over dx}\right)^2 - 2
c(\rho) {d^2 \rho \over d x^2} =0 \, ,$$
where $a(\rho) = (2 \sigma(\rho))^{-1}$ and $c(\rho)= D^2(\rho) \
a(\rho)$.
Multiplying the above equation by $d\rho(x) / dx$,  one obtains
after one integration 
\begin{eqnarray}
\label{eq: derivee}
D^2(\rho) \left( {d \rho \over dx} \right)^2 = q^2 ( 1 + 2 K
\sigma(\rho)) \,  ,
\end{eqnarray}
where $K$ is a constant of integration.

To proceed further one needs to determine the sign of ${d \rho \over dx}$.
The simplest case is when $\rho(x)$ is monotone, and this happens when $q$ is close enough to its average value for $\rho_a \neq \rho_b$ (this corresponds to values of $K$ small enough for the right hand side  of (\ref{eq: derivee}) not to vanish).
The investigation of this regime is enough to determine all the cumulants.
If for example  $\rho_a > \rho_b$, 
 the optimal $\rho(x)$ is decreasing  for small $K$ 
\begin{equation}
{d \rho \over dx} = - {q \over D(\rho)} \sqrt{ 1 + 2 K \sigma(\rho)}
\, ,
\label{drho}
\end{equation}
and this leads to the following expression for $G$:
\begin{equation}
G =  q \int_{\rho_b}^{\rho_a} {D(\rho) \over
\sigma(\rho)} \left[ 1 - {1 +  K \sigma(\rho) \over \sqrt{1 + 2 K \sigma(\rho)} }\right] 
d \rho \, ,
\label{G}
\end{equation}
where the constant $K$ is determined by:
\begin{eqnarray}
\label{eq: K}
q= \int_{\rho_b}^{\rho_a} d \rho { D(\rho) \over \sqrt{1 + 2 K
\sigma(\rho)}} \, .
\end{eqnarray}
One can then  show that
$${\partial G \over \partial q} = {G \over q} +K q  =
\int_{\rho_b}^{\rho_a} d \rho { D(\rho) \over \sigma(\rho)} \left[ 1 - {1 \over \sqrt{1 + 2 K
\sigma(\rho)}}\right]$$
where the derivative is taken keeping $\rho_a$ and $\rho_b$ fixed, and using the fact that 
$\mu_N = N^{-1} \max_q [ \lambda q + G(q,\rho_a,\rho_b) ]$, 
one obtains (\ref{mu},\ref{lambda}). 

When  the optimal $\rho(x)$ is no longer monotone, i.e. $K$ is negative
enough for the right hand side of (\ref{eq: derivee}) to vanish, the
expressions (\ref{mu},\ref{lambda},\ref{G},\ref{eq: K})  of $\mu_N,\lambda,G,q$ are modified. We checked that their new expressions are simply the analytic continuations of (\ref{mu},\ref{lambda},\ref{G},\ref{eq: K}).

In general when the system is in equilibrium ($\rho_a=\rho_b=\rho$) the fluctuations given by
(\ref{eq: fonctionnelle}) are non Gaussian. However when
$\rho_a=\rho_b=\rho^*$ where $\rho^*$ is the density for which
$\sigma(\rho)$ is maximum,  the optimal $\rho(x)$ in (\ref{eq: fonctionnelle})
 satisfies $\rho'(x)=0$ and the fluctuations become Gaussian
($G(q,\rho^*,\rho^*)= - q^2/(2 \sigma(\rho^*))$) in agreement with the 
conjecture made  in \cite{DDR}
for a specific model, the symmetric simple exclusion process.

It is also  easy to check from (\ref{eq: fonctionnelle}) that the optimal profile $\rho(x)$
is the same for $q$ and $-q$. This implies that
$$ G(-q,\rho_a,\rho_b)= G(q,\rho_a,\rho_b)  - 2 q \int_{\rho_b}^{\rho_a}
{ D(\rho) \over \sigma(\rho)} d \rho $$
which is the  Gallavotti-Cohen relation \cite{GC,LS,DDR}.

\begin{figure}[h!]
  \begin{center}
    \leavevmode
    \epsfxsize=7.8cm
    \epsfbox{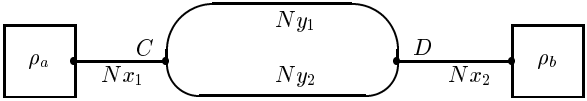}
    \caption[]
    {
The system connecting the two reservoirs contains a loop with two arms of
unequal lengths.}
  \label{FIG}
  \end{center}
\end{figure}

Consider now a system 
 composed of 4 parts
as in Figure \ref{FIG}.  
The left reservoir is connected to  $C$ by a  chain of 
length $Nx_1$. Between  $C$ and $D$ there is a loop made of
two  chains in parallel of lengths $Ny_1$ and $Ny_2$ and  
$D$ is connected  to the right reservoir by a chain of length $Nx_2$.
According to the {\it additivity principle}, one should have
\begin{eqnarray*}
&G_{\rm loop}(q,\rho_a,\rho_b) = \max_{\rho_c,\rho_d,q'} \left[
{G(qx_1,\rho_a,\rho_c) \over x_1} +
 \right.\\
&  \left.
 {G(q'y_1,\rho_c,\rho_d) \over y_1}
+ {G((q-q')y_2,\rho_c,\rho_d) \over y_2} +
{G(qx_2,\rho_d,\rho_b) \over x_2} \right] 
\end{eqnarray*}
The optimum is achieved when $q'( y_1+ y_2)=q y_2 $,
 thus
$$G_{\rm loop}(q,\rho_a,\rho_b) = G(qu,\rho_a,\rho_b)  /  u \, ,$$
with
$u = x_1 + (y_1^{-1}+y_2^{-1})^{-1} + x_2$.
So the current fluctuations for the  system with a loop are the same as for a linear system with a length
given by Kirchoff's law for the addition of resistors.


We consider now two specific examples of stochastic dynamics on a $1d$ 
lattice, the symmetric simple exclusion process (SSEP) and the zero 
range process (ZRP).
The number of particles at site $i \in \{0,N\}$ is denoted by 
$\eta_i$.

In the SSEP,  each site is is either empty or occupied by a single particle 
($\eta_i=0$ or $1$) and  each particle attempts 
to jump to its right or to its left at rate 1 if there is no other particle
at the corresponding neighboring site.
Let $Q^i_t$ be the integrated current through  bond $(i, i+1)$ during time $t$.
As in the steady state the integrated current is independent of the bond
and  $\partial_t \langle Q^i_t \rangle= \langle \eta_i - \eta_{i+1}
\rangle$, 
$$
N \partial_t \langle Q_t \rangle =  \sum_i \langle \eta_i - \eta_{i+1}
\rangle = \rho_a - \rho_b \, ,  $$
thus $ D(\rho) = 1 $. 
As $ N^2 \partial_t \langle (Q_t)^2 \rangle 
=
\partial_t  \langle \Big( \sum_i Q^i_t \Big)^2 \rangle$, we write
\begin{eqnarray*}
\sum_{i,j} \partial_t  \langle Q^i_t \, Q^j_t  \rangle
&=  \sum_{j,i} \langle Q^j_t  \big(\eta_i (1 -\eta_{i+1})
- \eta_{i+1} (1 -\eta_i) \big) \rangle \\
& + \sum_{i} \langle  \eta_i (1 -\eta_{i+1})\rangle
+ \langle  \eta_{i+1} (1 -\eta_i) \rangle \\
=  \sum_{j,i} & \langle Q^j_t  \big(\eta_i - \eta_{i+1} \big) \rangle 
+ 2 \sum_{i} \langle \eta_i (1 -\eta_{i+1}) \rangle \, .
\end{eqnarray*}
The first term simplifies
\begin{eqnarray*}
\sum_{j,i} \langle Q^j_t  \big(\eta_i - \eta_{i+1} \big) \rangle 
= \langle \big( \sum_j Q^j_t \big) \eta_0   \rangle 
- \langle \big( \sum_j Q^j_t \big) \eta_N \rangle \, .
\end{eqnarray*}
and it vanishes for  $\rho = \rho_a = \rho_b$. 
For $\rho_a=\rho_b$, the stationnary measure is product so 
that $\sigma(\rho) = 2 \rho( 1- \rho)$
according to  (\ref{eq: conductivite}).
The cumulants derived in \cite{DDR} as well as the 
expression  conjectured for $\mu_N$ 
\begin{eqnarray}
\label{eq: sol SSEP}
\mu_N(\lambda) = - N^{-1}  [\sin^{-1}(\sqrt{-\omega})]^2 \,,  \ \ \ {\rm
for } \  \omega \leq 0.
\end{eqnarray}
 where $\omega= (1 - e^{-\lambda}) ( e^\lambda \rho_a - \rho_b -
(e^\lambda-1) \rho_a \rho_b)$
can be recovered
from (\ref{mu},\ref{lambda}).
This can be seen  by noticing that the optimal profile solution of
(\ref{drho})  is
$$ \rho(x) = {1 \over 2} \left(1 + 
{\sin \big(2 \big(\theta_a + (\theta_b-\theta_a) x \big) \big)
\over \sin (2 f)} \right) \,, $$
where the parameters  $f,
 \theta_a,\theta_b$ are fixed by 
$K=\tan^2( 2 f),  \rho(0)=\rho_a, \rho(1)=\rho_b$.
In terms of these parameters, $\lambda$ and $\mu_N$ take the form
$$\lambda = \log \left[ {\cos(f + \theta_a) \ \cos(f - \theta_b) \over
\cos(f - \theta_a) \ \cos(f + \theta_b) } \right], \quad
\mu_N= - (\theta_a - \theta_b)^2 \, .$$

For the ZRP the number of particles on each site can be arbitrary 
and the jump rate $\Phi(\eta_i)$ from site $i$  to each of its neighbors is an increasing function of  
 the number of particles  $\eta_i$ at this site.
We choose $\Phi(0)= 0$. We have
$$
 \sum_i \partial_t \langle Q^i_t \rangle = 
\sum_i \langle \Phi(\eta_i) - \Phi(\eta_{i+1}) \rangle 
= \Psi(\rho_a) -  \Psi(\rho_b) \, ,
$$
where the expectation of $\Phi$ under the stationary measure at density $\rho$ 
is denoted by $\Psi(\rho)$.
We also have
\begin{eqnarray*}
N^2 \partial_t \langle (Q_t)^2 \rangle 
=  
\sum_{j,i} \langle Q^j_t  \big( \Phi(\eta_i) - \Phi(\eta_{i+1}) \big) \rangle 
+ 2 \sum_{i} \langle \Phi(\eta_i) \rangle  \, .
\end{eqnarray*}
As for the SSEP the 1st term  in the rhs of the above equation vanishes 
when $\rho_a=\rho_b$.
We finally obtain 
 $D(\rho) = \sigma^\prime(\rho)/2$  and
$\sigma(\rho) = 2 \Psi(\rho)$ according to 
 (\ref{eq: diffusivite},\ref{eq: conductivite}).
Therefore $$\mu_N(\lambda) = (1 - e^{-\lambda}) \big(e^{\lambda} \sigma(\rho_a) -
\sigma(\rho_b)\big)/2N \  .$$
This generalizes  the case of non-interacting
particles  for which   $\sigma(\rho)=2\rho$.
The optimal profile is obtained by the change of  variables
$$\sigma(\rho(x)) ={(\theta_a + (\theta_b - \theta_a)x)^2 -1 \over 2 K}\,,$$
where $\theta_a,\theta_b$ are fixed by 
$\rho(0)=\rho_a$
and $\rho(1)=\rho_b$.
In particular, the expression of $\mu_N$ follows from
$$
\lambda= \log\left( {1 + \theta_b \over 1 + \theta_a }\right), \quad
\mu_N(\lambda) = - {(\theta_a - \theta_b)^2 \over 4K} \, .
$$

The {\it additivity principle} (\ref{ld},\ref{add1}) formulated here
and its  variational expression (\ref{eq: fonctionnelle}) can be
derived (work in progress) from the hydrodynamic large deviation theory
\cite{KOV,L,HS}.
This theory was extended recently by 
Bertini et al. \cite{BDGJL}
who could  calculate the density large deviation functional of the
steady state as the optimal cost per unit time  for a space/time density fluctuation.
For diffusive systems, the exponential cost of observing an atypical space/time
density profile over a time $t$ can be estimated  by a functional
depending only on $D(\rho),\sigma(\rho)$ and on the density 
$\{\rho(x,s)\}_{x \in [0,1],0\leq s \leq t}$ (see eg \cite{KOV,L,HS}).
The optimal strategy to create a fluctuation of the current $Q_t = q t$ over a very 
long time $t$ is to create  a  fixed  density profile $\rho(x)$ in order to facilitate
the deviation of the current and 
 (\ref{eq: fonctionnelle}) can be understood as 
the cost for maintaining this   atypical density profile.
The optimal profile controling  here the current fluctuations is  time
independent, in contrast to the one which controls the steady state
density fluctuations that Bertini et al.  \cite{BDGJL} had to
calculate. This is why our task here was easier and the additivity principle (\ref{add1}) is  simpler   than
the one obtained in \cite{DLS1} for the steady state fluctuations of the
density.

It would be interesting to see whether 
the Bertini et al.  macroscopic fluctuation theory satisfies
a generalized additivity principle for  time-dependent densities in 
the reservoirs. 
Other interesting extensions of the present work include the study of the
effect of  asymmetry in the bulk dynamics (i.e. of a field which  favors
jumps of particles from left to right) \cite{krug,popkov,BECE,ED,evans}
or the analysis of more complex networks, in particular of systems in
contact with three or more reservoirs \cite{Maes,EZ,BB,SL,NB}.

Of course, a very challenging issue would be to see whether the 
additivity principle could be valid for some mechanical systems
satisfying (\ref{eq: diffusivite},\ref{eq: conductivite}) 
without the need of an intrinsic source of noise as the stochastic 
systems considered here.

We would like to thank B. Dou{\c c}ot, G. Giacomin, J. L. Lebowitz, 
P.-E. Roche and F. van Wijland for helpful  discussions.
T.B. acknowledges the hospitality of Rutgers University 
and the support by NSF Grant DMR 01-279-26.

\end{document}